\documentstyle[aps,graphics]{revtex}
\begin{document} 

\title{Noiseless phase quadrature amplification via electro-optic 
feed-forward}

\author{Ben C.  Buchler, Elanor H.  Huntington and Timothy C. Ralph}

\address{Department of Physics, Faculty of Science\\
The Australian National University\\
Australian Capital Territory 0200, Australia}

\maketitle
\begin{abstract}
Theoretical results are presented which show that noiseless phase 
quadrature amplification is possible, and limited experimentally only 
by the efficiency of the phase detection system. Experimental results 
obtained using a Nd:YAG laser show a signal gain of $10dB$ and a signal 
transfer ratio of $T_{s}=0.9$. This result easily exceeds the standard 
quantum limit for signal transfer.  The results also explicitly 
demonstrate the phase sensitive nature of the amplification process.
\end{abstract}

\begin{center}
Submitted to Phys. Rev. A	
\end{center}

\section{Introduction}
The transmission of small optical signals carried by coherent light is 
fraught with difficulty. Attenuation of the optical power causes
degradation of the signal-to-noise ratio as the signal recedes into the quantum 
noise of the beam. A simple method to rectify this problem would be to 
amplify the signal. However this is not a trivial exercise as phase 
insensitive amplifiers, such as laser amplifiers, introduce excess 
noise \cite{haus62}. In the limit of high gain, the noise penalty for 
a phase insensitive amplifier is \textit{3dB}, ie a halving of the 
signal to noise ratio. This is known as the standard quantum limit 
(\textit{SQL}) for phase insensitive amplifiers. To overcome this 
problem, phase sensitive amplification is required \cite{caves81}. 
This can entail a non-linear process, as in the case of the optical 
parametric amplifier, where one observable is amplified while the 
conjugate is de-amplified. \cite{lev93}.

A far simpler method has been demonstrated by Lam \textit{et al.} 
\cite{lam97} using positive electro-optic feed-forward. In this scheme, a part 
of the input beam is tapped off using a beamsplitter and detected. The 
signal is then added back to a modulator further downstream. Using 
this method with amplitude signals, Lam \textit{et al.} achieved a 
signal transfer ratio of $T_{s}=0.88$.

In this work we use an analogous feed-forward network to show 
better-than-\textit{SQL} amplification of phase quadrature signals. To our 
knowledge, this is the first demonstration of phase signal 
amplification superior to the amplifier \textit{SQL}.

In \textit{any} experiment which generates phase quadrature signals 
close to the quantum noise limit (\textit{QNL}) our system could be 
used to make this signal robust optical loss. Specific suggestions for 
the application of phase feed-forward include quantum non-demolition 
measurements \cite{ral97} and continuous variable teleportation 
\cite{brau,ral98}.

\section {Theory} 
\begin{figure}[h!]
\centering 
\resizebox{13cm}{!}{\includegraphics{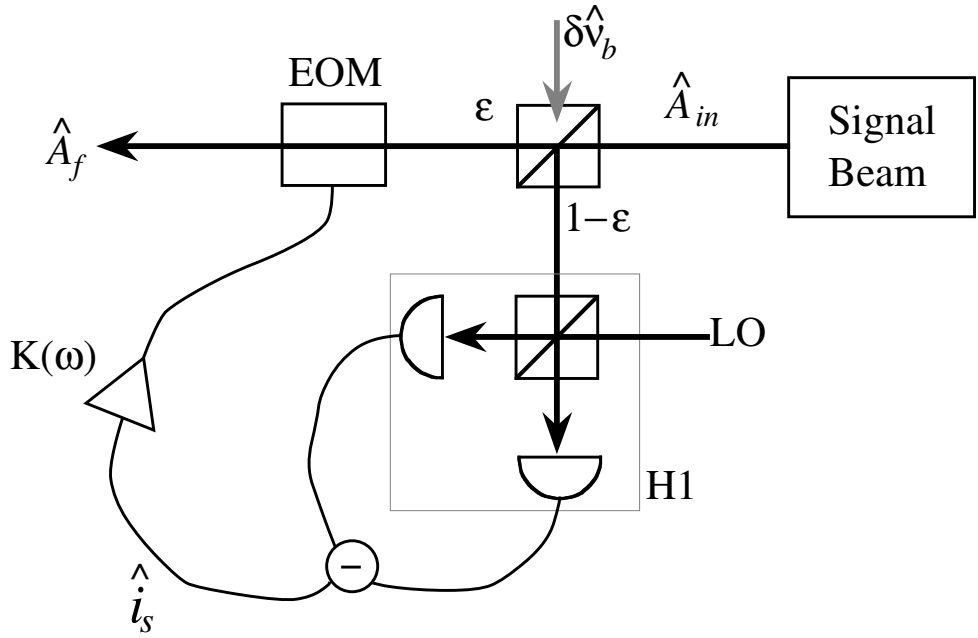}}
  \caption{A schematic diagram showing the components required for 
  noiseless phase quadrature amplification using phase feed-forward}
  \label{idea} 
\end{figure}

We will model the system shown in figure \ref{idea}. 
At the input to the system is a beam containing some phase signal, 
some of which is tapped off to the homodyne detection system 
(\textit{H1}) by a beamsplitter of transmissivity $\varepsilon$. The 
remaining light is passed through an electro-optic phase modulator, 
\textit{EOM}, which is controlled by the signal detected in 
\textit{H1}.

A theoretical model of the behaviour of this system will be generated 
using linearized operators. This is the same approach used previously 
in \cite{lam97} and \cite{ral97}. The time domain annihilation 
operator for the input beam to our system, $\hat{A}_{in}$, will be 
written as
\begin{equation}
	\hat{A}_{in}=\bar{A}_{in}+\delta \hat{A}_{in}
	\label{Ain}
\end{equation}
where $\bar{A}$ is the mean value of the field amplitude and $\delta 
\hat{A}$ is the time dependent component of the field with an 
expectation value of 0. Embedded in this time dependant component is 
the signal information and quantum noise carried by the beam. After 
traversing the beam path through the beamsplitter and \textit{EOM}, 
the field operator $\hat{A}_{f}$ for the output can be written as
\begin{equation}
\hat{A}_{f}=\sqrt{\varepsilon} \bar{A}_{in} + \sqrt{\varepsilon}\delta 
\hat{A}_{in}-\sqrt{1-\varepsilon} \delta \hat{\nu}_{b}+\imath \delta \hat{r}
	\label{Af}
\end{equation}
where $\delta \hat{\nu}_{b}$ is the vacuum input due to the beamsplitter and 
$\delta \hat{r}$ is a modulation imposed on the beam by the \textit{EOM}. 
The operator $\delta \hat{r}$ will be a function of fluctuations 
detected in the homodyne detection system, \textit{H1}.

A homodyne system, such as that shown in figure \ref{idea}, measures a 
particular quadrature amplitude of a low power beam by mixing it with a
much more intense local oscillator beam on a 50/50 beamsplitter 
\cite{homodyne}. The phase difference between the local oscillator and 
the signal beam, $\theta$, determines which quadrature of the signal 
beam is measured. The output of the system is the difference of the 
photocurrents from the two photo-diodes. The form of this current may 
be calculated by finding the difference of the photon number operators 
(i.e. $\hat{A}^{\dagger}\hat{A}$) for the fields incident on each of 
the detectors. To simplify matters we will linearize the equations by 
neglecting all terms greater than first order in the fluctuation 
operators. The form of the subtracted photocurrent is found to be
\begin{eqnarray}
\hat{i}_{s}&=& 2\sqrt{\eta_{d}\eta_{h}} 
\bar{A}_{s} \cos \theta + \sqrt{\eta_{h} 
\eta_{d}(1-\varepsilon)}\delta 
\hat{X}^{\theta}_{A_{in}} \nonumber\\
&&{}+\sqrt{\eta_{d}}\left(\sqrt{\eta_{h} 
\varepsilon}\delta \hat{X}^{\theta}_{{\nu}_{b}} + \sqrt{1-\eta_{h}} \delta 
\hat{X}^{\theta}_{{\nu}_{h}}\right) \nonumber\\
&&{}+ \frac{\sqrt{1-\eta_{d}}}{\sqrt{2}}(\delta 
\hat{X}_{{\nu}_{d1}}+\delta \hat{X}_{{\nu}_{d2}}),
\label{is}
\end{eqnarray}
where we have defined the general quadrature of an operator $\hat{z}$ to be 
\begin{equation}
\delta \hat{X}^{\theta}_{z}=e^{-\imath \theta} \hat{z}+e^{\imath 
\theta} \hat{z}^{\dagger}.
\end{equation}
for convenience of notation, the amplitude quadrature ($\theta=0$) 
will be written with no superscript, while the phase quadrature 
($\theta=\pi/2$) will be written as $\hat{X}^{-}_{z}$. The quantum 
noise terms $\delta \hat{X}_{{\nu}_{d1,d2}}$ result from the 
efficiency $\eta_{d}$ of the two photodetectors in the homodyne system. 
The mode-matching efficiency of the homodyne system $\eta_{h}$ also 
gives rise to a source of quantum noise, namely $\delta 
\hat{X}^{\theta}_{h}$. The mode-matching efficiency, $\eta_{h}$, is 
given by the square of the fringe visibility in the homodyne system 
\cite{hansbook} i.e.
\begin{equation}
\mathrm{Homodyne\ efficiency}=\eta_{h}= 
\left(\frac{I_{max}-I_{min}}{I_{max}+I_{min}}\right)^{2}
\end{equation}
where $I_{min/max}$ is the minimum/maximum power of the output of one 
arm of the homodyne system as measured with equal power in the signal 
and local oscillator arms. To measure the phase quadrature with 
homodyne detection, we require that $\theta=\pi/2$. This means that 
the \textit{DC} component of the current in equation \ref{is} is zero. 
Experimentally, this provides a means for locking the homodyne system 
to the phase quadrature.

The form of the modulation $\delta \hat{r}$ imposed by the \textit{EOM} 
will be given by a convolution of $\delta \hat{i}_{s}$, the time 
dependent part of the current $\hat{i}_{s}$, and the time response of 
the feed-forward electronics, $k(t)$. We may therefore write
\begin{equation}
\delta \hat{r}(t)=\int_{0}^{t}k(u) \delta \hat{i}_{s}(t-u)du.
\label{r}
\end{equation}
Combining equations \ref{Af}, \ref{is} and \ref{r}, as well as 
choosing the phase of the homodyne detection system to be $\pi/2$ so 
that the phase quadrature is being fed-forward, the final form of the 
fluctuations in the output beam of the \textit{EOM} will be given by
\begin{eqnarray}
\lefteqn{\delta \hat{A}_{f}=\sqrt{\varepsilon}\,\delta 
\hat{A}_{in}-\sqrt{1-\varepsilon} \,\delta 
\hat{\nu}_{b}\label{Af'}} \\
&&{}+\imath \int_{0}^{t}k(u)\Big[\sqrt{\eta_{h} 
\eta_{d}(1-\varepsilon)}\,\delta \hat{X}^{-}_{A_{in}}(t-u) \nonumber \\
	&&{}+\sqrt{\eta_{d}}\left(\sqrt{\eta_{h} 
{\varepsilon}}\,\delta \hat{X}^{-}_{{\nu}_{b}}(t-u) + \sqrt{1-\eta_{h}} \,\delta 
\hat{X}^{-}_{{\nu}_{h}}(t-u)\right) \nonumber \\ 
&&{}+\frac{\sqrt{1-\eta_{d}}}{\sqrt{2}}\left(\delta 
\hat{X}_{{\nu}_{d1}}(t-u)+\delta 
\hat{X}_{{\nu}_{d2}}(t-u)\right)\Big]du.\nonumber 
\end{eqnarray}
Considering the general quadrature, $\phi$, of $\delta \hat{A}_{f}$ and taking 
the Fourier transform of the operators we obtain
\begin{eqnarray}
\delta \tilde{X}_{A_{f}}^{\phi}&=&\left(K(\omega)\sin\phi \sqrt{\eta_{h} 
\eta_{d}(1-\varepsilon)}\,\delta 
\tilde{X}_{A_{in}}^{-}+\sqrt{\varepsilon}\,\delta 
\tilde{X}_{A_{in}}^{\phi}\right) \nonumber \\
&&{}+ \left(K(\omega)\sin\phi \sqrt{\eta_{d}\eta_{h} {\varepsilon}}\,\delta 
\tilde{X}_{\nu_{b}}^{-}- \sqrt{1-\varepsilon}\,\delta 
\tilde{X}_{\nu_{b}}^{\phi}\right) \nonumber \\
&&{} + K(\omega)\sin\phi \big[\sqrt{\eta_{d}}\sqrt{1-\eta_{h}} \,\delta 
\tilde{X}^{-}_{{\nu}_{h}} \label{XAf} \\
&&{}+ \frac{\sqrt{1-\eta_{d}}}{\sqrt{2}}(\delta 
\tilde{X}_{{\nu}_{d1}}+\delta \tilde{X}_{{\nu}_{d2}})\big],\nonumber
\end{eqnarray}
where $\tilde{X}_{z}$ is the Fourier transform of $\hat{X}_{z}$. The 
spectrum of $\hat{A}_{f}$, normalized to the quantum noise limit 
(\textit{QNL}), can now be obtained for arbitrary quadrature phase 
angle by evaluating $V_{Af}^{\phi}=<|\delta 
\tilde{X}_{A_{f}}^{\phi}|^{2}>$. Using the result that the spectra of 
the quantum noise sources is just $1$, we obtain
\begin{eqnarray}
V_{Af}^{\phi} &=& \sqrt{\frac{{V^{-}_{A_{in}}}^{2}}{\sin^{2} \alpha + 
{V^{-}_{A_{in}}}^{2}\cos^{2}\alpha}} \times \nonumber \\
&&\left(\varepsilon \cos^{2}\phi + \left| 
\sqrt{\varepsilon } + K(\omega) \sqrt{\eta _{h} \eta _{d} (1 - 
\varepsilon)} \right| ^{2}\sin^{2}\phi\right) 
 \nonumber \\
 & &+\left|K(\omega) \sqrt{\eta _{d} \eta _{h} \varepsilon} - \sqrt{1 - 
 \varepsilon }\right|^{2}\sin^{2} \phi \label{VAf} \\
 &&{}+ (1 - \varepsilon ) 
 \cos^{2} \phi + \left|K(\omega)\right|^{2} (1 - \eta _{d} \eta 
 _{h})\sin^{2}\phi \nonumber
\end{eqnarray}
where the angle $\alpha$ is given by 
\[\tan 
\alpha=\left(1+\frac{K(\omega)\sqrt{\eta_{h}\eta_{d} 
(1-\varepsilon)}}{\sqrt{\varepsilon}}\right)\tan \phi.\] 
It has also been assumed that the amplitude noise of the input beam, 
$\hat{A}_{in}$, is at the quantum limit. 
To investigate the action of the feed-forward system on the phase 
quadrature, we take $\phi=\pi/2$ so that
\begin{eqnarray}
	V_{Af}^{-} &=& \left| 
\sqrt{\varepsilon } + K(\omega) \sqrt{\eta _{h} \eta _{d} (1 - 
\varepsilon)} \right| ^{2} V_{A_{in}}^{-} \nonumber\\
&&{}+\left|K(\omega) \sqrt{\eta _{d} \eta _{h} \varepsilon} - \sqrt{1 - 
\varepsilon }\right|^{2} \nonumber \\
&&{} + \left|K(\omega)\right|^{2} (1 - \eta _{d} \eta _{h})
\label{VAf-}
\end{eqnarray}
From equation \ref{VAf-} it is clear that a value of the electronic 
gain may be chosen such that the second term in equation \ref{VAf-} 
becomes 0. Physically this may explained by the division of the 
beamsplitter vacuum fluctuations $\delta \hat{\nu}_{b}$. The component 
of $\delta \hat{\nu}_{b}$ imposed on the signal beam (i.e. the beam 
which goes through the \textit{EOM}) is anti-correlated with the 
component of $\delta \hat{\nu}_{b}$ imposed on the light passed to the 
homodyne system \textit{H1}. The result is that that for a unique 
value of \textit{positive} feed-forward gain, the \textit{EOM} will be 
driven such that the signal it imposes exactly cancels the 
beamsplitter vacuum fluctuations originally introduced to the signal 
beam. If we assume ideal homodyne and detector efficiency, the value 
of the gain required for cancellation is 
\begin{equation}
	K(\omega)=\sqrt{\frac{1-\varepsilon}{\varepsilon}}.
	\label{gaino}
\end{equation}
With this gain, the phase noise of the beam at the output of the 
feed-forward modulator is given by
\begin{equation}
V_{A_{f}}^{-}=\frac{V_{A_{in}}^{-}}{\varepsilon}
\label{VAfo}
\end{equation}
Assuming perfect in-loop detection, the system can now been seen to 
behave as a noiseless amplifier with signal amplification of 
$\varepsilon^{-1}$.

When detection losses are considered, matters become a little more 
complex. The optimum gain is no longer that which gives total 
cancellation of the beamsplitter vacuum noise. This is because the 
extra noise due to detection scales with feed-forward gain. The result 
is that the optimum gain level is less than that found for ideal 
detection. To evaluate the performance of this system with imperfect 
detection we define a signal transfer ratio $T_{s}$. This is given by 
the ratio of the signal-to-noise at the output of the system to the 
signal-to-noise at the input of the system, i.e.
\begin{equation}
	T_{s}=\frac{SNR_{out}}{SNR_{in}} \label{Ts}
\end{equation}
The optimum value of $T_{s}$ occurs at a gain of
\begin{equation}
	K(\omega)=\sqrt{\frac{\eta_{h}\eta_{d}(1-\varepsilon)}{\varepsilon}},
	\label{gain}
\end{equation}
for which
\begin{equation}
	T_{s}=\varepsilon(1-\eta_{h}\eta_{d})+\eta_{h}\eta_{d}. \label{Tso}
\end{equation}
We note that with ideal detection we obtain $T_{s}=1$ indicating 
perfect signal transfer.

When using a feed-forward system as a signal amplifier the optical 
power of the signal beam is attenuated by a factor of $\varepsilon$. 
This power loss can be overcome by using an injection locking system 
\cite{forlock}. The final output beam can then be an optical state 
with identical power to the original state, but with an amplified 
signal.

\section{The Experiment}
\begin{figure}[h!]
\centering 
\resizebox{13cm}{!}{\includegraphics{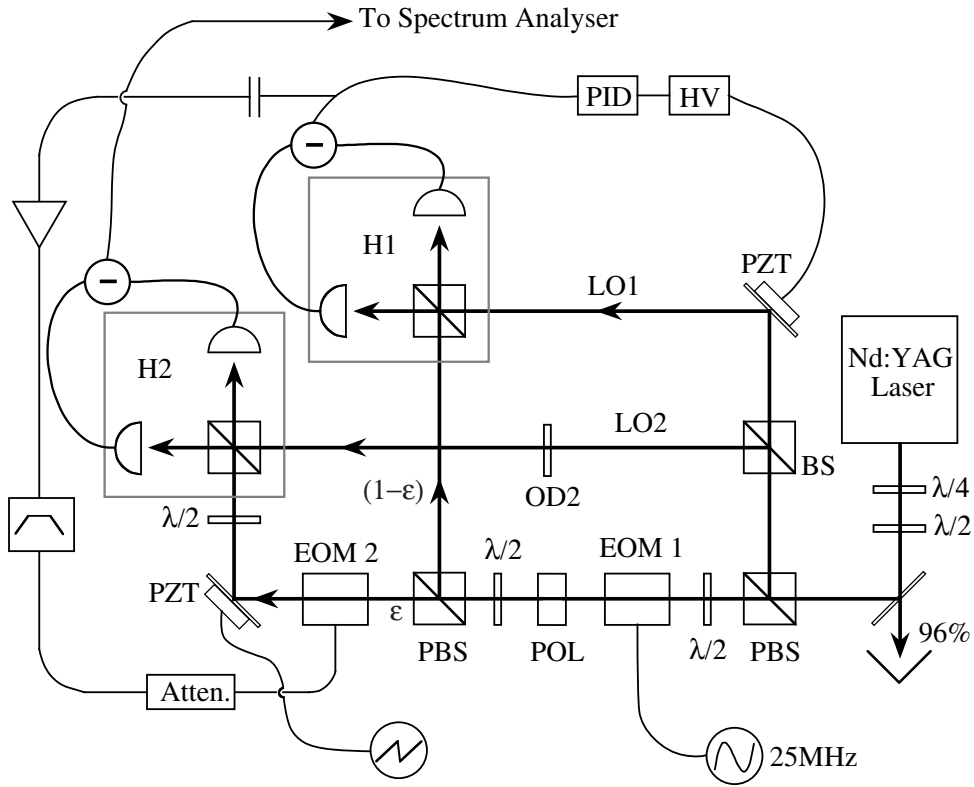}}
  \caption{The layout of our phase feed-forward experiment.}
  \label{setup} 
\end{figure}

An experiment was set up as shown in figure \ref{setup}. The light 
source was a single mode \textit{Nd:YAG} laser pumped by a diode laser 
array. The output power of the laser was \textit{300mW} at 
\textit{1064nm}. Most of the light was dumped at the 96\% 
beamsplitter. Of the remaining light, 99.75\%, was tapped off for the 
local oscillator beams \textit{LO1} and \textit{LO2}.  The remaining 
light formed the signal beam for the phase feed-forward.  The 
electronics used for the feed forward consisted of two high gain 
amplifiers and a bandpass filter to ensure sufficient gain at the 
signal frequency.  A variable electronic attenuator was used to adjust 
the overall gain of the loop.

A phase signal was imposed on the beam using the phase modulator 
\textit{EOM 1}.  A frequency of $25MHz$ was chosen because the laser 
was at the \textit{QNL} at this frequency.  The signal was measured 
using the homodyne detection systems \textit{H1} and \textit{H2}.  The 
homodyne detector \textit{H1} consisted of two identical low noise 
detectors with a quantum efficiency of $\eta_{d}=0.91 \pm 0.02$ 
\cite{det}.  The homodyne 
efficiency was measured to be $\eta_{h}=0.94\pm 0.02$.  System \textit{H2} 
contained similar detectors also with a quantum efficiency of 
$\eta_{d}=0.91 \pm 0.02 $.  The homodyne efficiency of \textit{H2} was 
$0.88 \pm 0.02$.

The homodyne system \textit{H1} was locked to the phase quadrature 
using the DC voltage from the subtraction of the photocurrents of the 
homodyne system as an error signal. As discussed previously in relation 
to equation \ref{is}, when the DC voltage from $H1$ is 0, the 
detection system measures the phase quadrature.

\begin{figure}[h!]
\centering 
\resizebox{13cm}{!}{\includegraphics{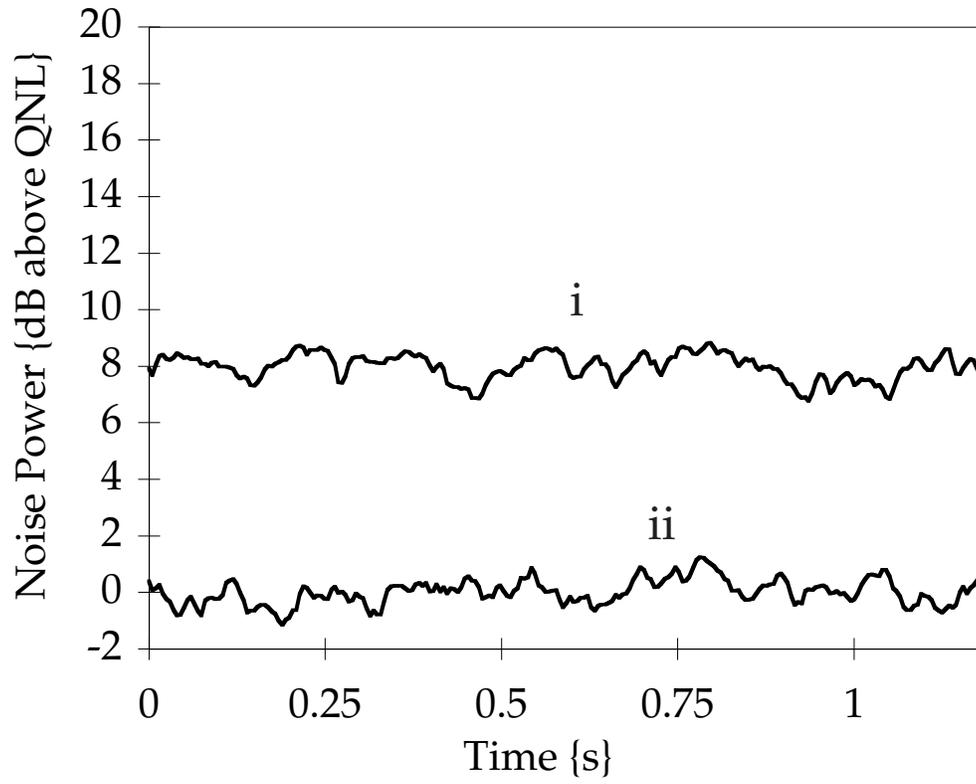}}
  \caption{Trace 
  \textit{i} shows a plot of the input phase signal at \textit{25MHz}. 
  This measurement was made with 100\% of the signal beam directed into 
  the homodyne detector \textit{H1} and is therefore a measurement of 
  $V_{A_{in}}$. Trace \textit{ii} show the quantum noise. The spectrum 
  analyser was set on a resolution bandwidth of \textit{100Hz} and video 
  bandwidth \textit{3Hz}.}
  \label{8020input} 
\end{figure}

The phase modulation imposed by \textit{EOM 1} is shown in figure 
\ref{8020input}, as measured using a spectrum analyzer 
(\textit{Hewlett-Packard 3589 A}). The upper trace shows the signal 
level with 100\% of the signal beam directed into the locked homodyne system 
\textit{H1}. The signal is observed to be $8.0\pm0.4dB$ above the 
quantum noise, shown at $0dB$.  Taking detection efficiency into 
account means that the inferred signal level is $8.6\pm0.4dB$ above 
the \textit{QNL}.

\begin{figure}[h!]
\centering 
\resizebox{13cm}{!}{\includegraphics{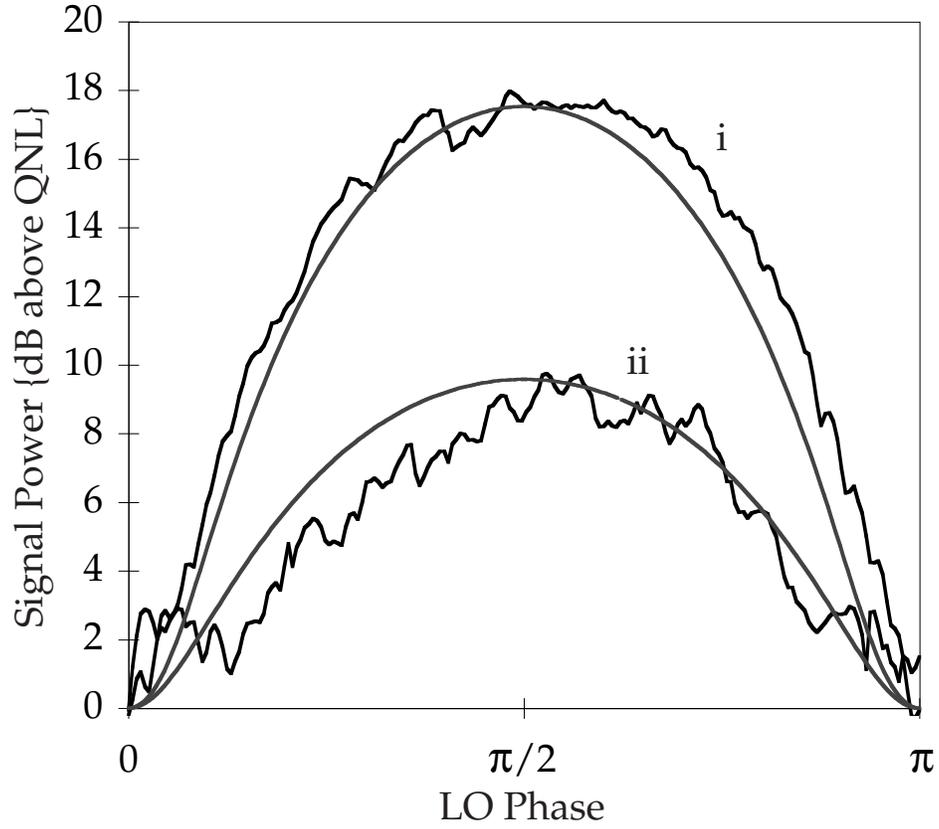}}
\caption{Trace 
\textit{i} shows a plot of the phase signal measured at \textit{25MHz} 
with 80\% of the light directed into \textit{H1} and the feed-forward 
optimised. Trace \textit{ii} show the noise level at 
\textit{25.0003MHz}. The theoretical fit to the data was obtained using 
equation \ref{VAf}. The spectrum analyser was set on a resolution 
bandwidth of \textit{100Hz} and video bandwidth \textit{3Hz}. This 
plot is an average of 3 sweeps.}
  \label{8020} 
\end{figure}

We then altered our system so that 80\% of the light was directed to 
\textit{H1} while the remaining 20\% was passed through \textit{EOM 2} 
to the homodyne detector \textit{H2}.  This corresponds to 
$\varepsilon=0.2$ in the above theory. The signal measured by 
\textit{H1} was used to drive \textit{EOM 2} through the feed-forward 
loop.  The result is shown in figure \ref{8020}.  Homodyne system 
\textit{H2} was swept and signals are therefore plotted as a function of the 
phase of \textit{LO2}.  Trace \textit{i} shows the signal power at 
$25MHz$ on the phase quadrature to be $17.6\pm 0.2dB$ above the 
\textit{QNL}.  The signal gain of the system is therefore $\approx 
10dB$.

The signal level measured after feed-forward must be compared to the 
noise observed at a frequency slightly removed from the signal 
frequency.  Trace \textit{ii} of figure \ref{8020} shows the phase 
quadrature noise power as measured at $25.0003MHz$ to be at $9.5\pm 
0.4 dB$ above the \textit{QNL}.  The difference between traces 
\textit{i} and \textit{ii} when the phase of \textit{LO2} is $\pi/2$ 
shows the fed-forward signal to be $8.1\pm 0.4dB$ above the noise 
floor.

A theoretical fit of this data may be obtained using 
equation \ref{VAf}. This is also shown in figure \ref{8020}. The only 
free parameter in this fit is the gain $K(\omega)$ which we 
take to be a constant $K$ at a fixed frequency of $25MHz$. The value 
of $K$ used to give this fit is 3.2, which is 70\% (or $2.4dB$) larger 
than the calculated optimum. However with high signal gain the system 
is insensitive to detuning from optimum gain \cite{lam97}.

Importantly, figure \ref{8020} demonstrates that the 
system is truly phase sensitive. The amplitude quadrature of the 
signal beam, as measured at phase angles of $0$ and $\pi$, is shown to 
be quantum noise limited.

From an operational point of view, the system is very successful. We 
measure an input signal of $8\pm0.4dB$ above the noise, and retrieve a 
signal $8.1\pm0.4dB$ above the noise. This new signal is almost $18dB$ 
above the \textit{QNL} instead of $8dB$ and therefore more immune to 
attenuation. To rigorously determine how our system rates as a 
noiseless amplifier, we employ the signal transfer ratio defined in 
equation \ref{Ts}. In this equation, $SNR_{in}$ is the inferred 
signal-to-noise ratio (\textit{not} the detected) of the input beam 
and $SNR_{out}$ is the inferred signal to noise of the output beam. 
Considering the homodyne and detection efficiency of \textit{H1}, we 
can infer an input signal-to-noise ratio of $SNR_{in}=6.2 \pm 0.7$. 
From the efficiency of the homodyne system \textit{H2} we find the 
inferred $SNR_{out}$ to be $5.6 \pm 0.6$, therefore giving 
$T_{s}=0.90\pm 0.14$. The uncertainty in this result is largely due to 
the swept operation of \textit{H2}. We expect that a locked homodyne 
system would allow more accurate determination of $T_{s}$.

Using equation \ref{Tso} we find that the detector and homodyne 
efficiencies of \textit{H1} limit the maximum achievable $T_{s}$ to 
$0.88$, which is in agreement with our experimental results.  The maximum 
transfer coefficient achievable with a \textit{PIA} with $10dB$ signal 
gain is 0.53, which our system easily exceeds.

\section{Conclusion}
We have developed a theoretical model of a phase feed-forward network 
and shown that it can behave as a noiseless phase quadrature 
amplifier in the limit of ideal phase homodyne phase signal detection.  
Experimental results with a Nd:YAG laser source demonstrate the 
practicality of the system.  A signal transfer ratio of 
$T_{s}=0.9\pm0.14$ with a signal gain of $10dB$ was measured.\\
The relative simplicity of this system and the demonstrated 
insensitivity to non-optimum gain make this system an ideal add-on to 
any experiment where close to \textit{QNL} phase signals need to be made robust 
to optical loss.  Finally, these results indicate that quantum 
non-demolition and continuous variable teleportation experiments 
relying on electro-optic control of the phase quadrature are feasible.

\acknowledgments The Authors would like to thank P.K. Lam and H.-A. 
Bachor for their valuable input. B.C. Buchler and E.H. Huntington are 
recipients of an Australian Postgraduate Awards. T.C. Ralph is an 
Australian Research Council Postdoctoral Fellow.  This work was 
supported by the Australian Research Council.

\newpage

\end{document}